\documentclass[reprint,amsmath,amssymb,aps,prd,showpacs]{revtex4-1}

\usepackage{graphicx}
\usepackage{dcolumn}
\usepackage{bm}
\usepackage{epsfig}
\usepackage{subfigure}
\usepackage{multirow}
\usepackage{amsmath}
\usepackage{hyperref}

\begin{document}

\title{Zeros in the magic neutrino mass matrix}

\author{Radha Raman Gautam}\email{gautamrrg@gmail.com}
 \affiliation{Department of Physics, Himachal Pradesh University,       Shimla -171005, INDIA.} 

\author{Sanjeev Kumar}\email{skverma@physics.du.ac.in}
 \affiliation{Department of Physics and Astrophysics, University of Delhi,\\ 
Delhi -110007, INDIA.} 

\date{\today}

\begin{abstract}
We study the phenomenological implications of the presence of two zeros 
in a magic neutrino mass matrix. We find that only two such patterns of 
the neutrino mass matrix are experimentally acceptable. 
We express all the neutrino observables as functions of 
one unknown phase $\phi$ and two known parameters 
$\Delta m^{2}_{12}$, $r=\Delta m^{2}_{12}/\Delta m^{2}_{23}$.
In particular, we find $\sin^2 \theta_{13}=(2/3)r/(1+r)$.
We also present a mass model  for the allowed textures 
based upon the group $A_{4}$ using type I+II see-saw mechanism.
\end{abstract}


\maketitle

\section{Introduction} 
\label{sec:intro}
The observation of non-zero reactor mixing angle 
\mbox{($\theta_{13}$) \cite{th13}}
was an important landmark in neutrino physics 
as it excluded the possibility of the $\mu-\tau$ symmetry \cite{mutau} 
as an exact symmetry of the neutrino mass matrix. 
Before this discovery, the tri-bimaximal (TBM) mixing \cite{hps} 
was an important feature in the neutrino mass models
as it correctly predicted the solar mixing angle ($\theta_{12}$)
and the atmospheric mixing angle ($\theta_{12}$). 
TBM mixing was thought to be a signature of 
some flavor symmetry in the Lagrangian that expresses itself 
as a residual symmetry in the neutrino mass matrix. However,
TBM mixing is in itself a combination of the following
two symmetries:
\begin{enumerate}
\item \textbf{Magic symmetry}. 
The sum of elements in any row or column 
of the neutrino mass matrix is identical \cite{magic}. 
\item \textbf{$\mu-\tau$ symmetry}. 
The neutrino mass matrix remains invariant 
after the interchange of the $\mu-\tau$ indices \cite{mutau}. 
\end{enumerate}

The neutrino mass matrix with $\mu-\tau$ symmetry implies
a vanishing value of $\theta_{13}$ and a maximal value of
$\theta_{23}$. Such a mass matrix has bi-maximal eigenvector
$v=(0~\frac{-1}{\sqrt{2}}~\frac{1}{\sqrt{2}})^T$.
After the measurement of a relatively large value of $\theta_{13}$,  
the neutrino mass matrix cannot have exact $\mu-\tau$ symmetry.
However, the neutrino mass matrix can still have the
magic symmetry. The corresponding mixing pattern,
called trimaximal (TM) mixing, has its middle column
identical to that of TBM mixing. The other two columns are
arbitrary within the unitarity constraints. 

TM mixing has been intensively studied in the literature \cite{tm2} and 
corresponding magic mass matrix has been realized in many neutrino
mass models \cite{tm2models}. 
The main limitation of the magic symmetry is that it is not much 
predictive. It predicts TM mixing that implies two sum-rules: 
one between the mixing angles $\theta_{12}$ and $\theta_{13}$ 
and another between the mixing angle $\theta_{23}$ and the CP violating 
Dirac phase $\delta$. To make the magic symmetry more predictive, 
we can combine it with some additional constraint. 
The simplest constraint that could combine with magic symmetry 
was the $\mu-\tau$ symmetry. But, the
observation of a non-vanishing $\theta_{13}$ has already ruled out 
this possibility. Another constraint can be the presence of zeros \cite{fgm, xingtz, tz} in the
magic neutrino mass matrix. In this work, we study this possibility. 

In Section \ref{sec:tm}, we highlight the salient features of 
TBM mixing pattern and review its relation with TM mixing. 
We identify the phenomenologically allowed textures of 
two zeros in the magic neutrino mass matrix in Section \ref{sec:2tm}. 
Then, we study the phenomenology of the viable textures 
in Section \ref{sec:pheno} and construct a mass model for them 
in Section \ref{sec:model}. Finally, we conclude in section \ref{sec:conclusions}.

\section{From TBM to TM mixing} 
\label{sec:tm}
TBM mixing matrix is
\begin{equation}
U_{TBM} = \left(
\begin{array}{ccc}
-\frac{\sqrt{2}}{\sqrt{3}} & \frac{1}{\sqrt{3}} & 0 \\ 
\frac{1}{\sqrt{6}} & \frac{1}{\sqrt{3}} & -\frac{1}{\sqrt{2}}\\ 
\frac{1}{\sqrt{6}} & \frac{1}{\sqrt{3}} & \frac{1}{\sqrt{2}} 
\end{array}
\right).
\end{equation} 
It is called the tri-bimaximal mixing matrix because 
the corresponding neutrino mass matrix
\begin{equation}
M_{TBM}=U_{TBM}^*M_{diag}U_{TBM}^{\dagger}
\end{equation}
has a trimaximal eigenvector $u=(\frac{1}{\sqrt{3}}~\frac{1}{\sqrt{3}}~\frac{1}{\sqrt{3}})^T$ and a bimaximal eigenvector 
$v=(0~\frac{-1}{\sqrt{2}}~\frac{1}{\sqrt{2}})^T$. 
Here, 
\begin{equation}\label{eq:diag}
M_{diag} =\left(
\begin{array}{ccc}
 m_{1} & 0 & 0 \\
 0 & e^{2 i \alpha } m_{2} & 0 \\
 0 & 0 & e^{2 i \beta } m_{3}
\end{array}
\right),
\end{equation}
where $m_1$, $m_2$, and $m_3$ are the three neutrino masses and
$\alpha$ and $\beta$ are two Majorana phases. TBM mass matrix 
$M_{TBM}$ is invariant under the transformations $G_u$ and $G_v$;
\textit{i.e.} $G_u^T M_{TBM} G_u=M_{TBM}$ and $G_v^T M_{TBM} G_v=M_{TBM}$
where $G_u=1-2uu^T$ and $G_v=1-2vv^T$. The transformation $G_u$ corresponds
to the magic symmetry and the transformation $G_v$ corresponds to the 
$\mu-\tau$ symmetry. A diagonal charged lepton mass matrix will be invariant under the transformation $F=$ diag$(1,\omega,\omega^2)$ where $\omega =$ exp$(\frac{2 \pi i}{3})$.  
In this way, the combined symmetry group generated by $G_u$, $G_v$ and $F$ is $S_4$ \cite{s4}. 
Such neutrino mass models, where some of the generators of a symmetry group are directly preserved in 
the lepton sector, are called direct models. Other set of models, 
where the observed symmetry in the lepton sector emerges 
accidentally, are called indirect models. For detailed discussion of 
this classification, see the references \cite{grimus, king}.

Since the neutrino oscillation experiments have measured a non-zero
$\theta_{13}$, the neutrino mass matrix $M_{\nu}$ cannot be 
invariant under the $\mu-\tau$ symmetry transformation $G_v$. 
However, $M_{\nu}$ can still be invariant under the magic 
symmetry transformation $G_u$. The magic symmetry is still allowed 
experimentally. The mixing matrix corresponding to the magic symmetry
is called trimaximal mixing (TM) and is given by
\begin{equation}\label{eq:tm}
U_{TM}=
\left(
\begin{array}{ccc}
 \sqrt{\frac{2}{3}} \cos \theta &
   \frac{1}{\sqrt{3}} & \sqrt{\frac{2}{3}}
   \sin \theta \\
 -\frac{\cos\theta}{\sqrt{6}}+\frac{e^{-i \phi} \sin
\theta}{\sqrt{2}} & \frac{1}{\sqrt{3}} &
   -\frac{\sin\theta}{\sqrt{6}}-\frac{e^{-i \phi} \cos\theta}{\sqrt{2}} \\
 -\frac{\cos\theta}{\sqrt{6}}-\frac{e^{-i \phi}
   \sin \theta}{\sqrt{2}} &
   \frac{1}{\sqrt{3}} & -\frac{\sin
   \theta}{\sqrt{6}}
+\frac{e^{-i \phi}
   \cos \theta}{\sqrt{2}}\end{array}
\right).
\end{equation}
Since, the middle column of TM mixing matrix is fixed 
to its TBM value ($u$),
the mixing matrix still has two free parameters ($\theta$ and $\phi$) 
after the unitarity constraints are taken into account. 
The corresponding neutrino mass matrix 
for TM mixing is called the magic mass matrix
and is given as 
\begin{equation}\label{eq:reco}
M_{magic}=U_{TM}^*M_{diag}U_{TM}^{\dagger}.
\end{equation}

\section{Zeros of the magic mass matrix} 
\label{sec:2tm}
In the basis where the charged lepton mass matrix is diagonal, 
there are seven mass matrices with two zeros \cite{fgm, xingtz} 
that are consistent with the current experimental data \cite{data}. 
They have been further classified in the
three classes which have been depicted in Table \ref{tab:2t}. 
When we combine the magic symmetry and the texture zeros, not all of 
the seven textures will be allowed.

\begin{table}[tb]
\begin{center}
\begin{tabular}{cc}
\hline
\hline
 Type  &        Constraining Equations         \\
 \hline
 $A_1$ &     $M_{ee}=0$, $M_{e\mu}=0$     \\
 $A_2$ &    $M_{ee}=0$, $M_{e\tau}=0$     \\
 $B_1$ &  $M_{e\tau}=0$, $M_{\mu\mu}=0$   \\
 $B_2$ &  $M_{e\mu}=0$, $M_{\tau\tau}=0$  \\
 $B_3$ &   $M_{e\mu}=0$, $M_{\mu\mu}=0$   \\
 $B_4$ & $M_{e\tau}=0$, $M_{\tau\tau}=0$  \\
 $C$   & $M_{\mu\mu}=0$, $M_{\tau\tau}=0$  \\
 \hline
 \hline
\end{tabular}
\end{center}
\caption{Seven allowed mass matrices with two zeros classified 
into three classes.}
\label{tab:2t}
\end{table}

A most general magic mass matrix can be parameterized as \cite{magic}
\begin{equation}\label{eq:magic}
M_{magic} =\left(
\begin{array}{ccc}
 a & b & c \\
 b & d & a+c-d \\
 c & a+c-d & b-c+d
\end{array}
\right).
\end{equation}

We can obtain the constraining equations for the
various allowed textures of two zeros in the magic mass matrix 
by substituting the respective constraints from Table \ref{tab:2t}
in Eq. (\ref{eq:magic}).

\subsection{Class \textbf{A}}

Magic neutrino mass matrices having textures 
$A_{1}$ and $A_{2}$ can be 
expressed as
\begin{equation}
 M^{A_1}_{magic} = \left(
\begin{array}{ccc}
0 & 0 & c \\ 0 & d & c-d\\ c & c-d & -c+d 
\end{array}
\right)
\end{equation}
and
\begin{equation}\label{eq:a2}
 M^{A_2}_{magic} = \left(
\begin{array}{ccc}
0 & b & 0 \\ b & d & -d \\ 0& -d & b+d
\end{array}
\right),
\end{equation}
respectively. The mass matrix for the magic $A_1$ texture can be 
rewritten as
\begin{equation}\label{eq:a1}
 M^{A_1}_{magic} = \left(
\begin{array}{ccc}
0 & 0 & c \\ 0 & c-\Delta & \Delta \\ c & \Delta & -\Delta 
\end{array}
\right),
\end{equation}
where $\Delta=c-d$. 
This redefinition brings our representations of the textures $A_1$ and $A_2$
at equal footing. These two magic zero textures are allowed 
experimentally for normal hierarchy.  
Their phenomenology is studied in the Section \ref{sec:pheno}.

\subsection{Class \textbf{B}}
The four magic mass matrices of class \textbf{B} are 
\begin{equation}
M^{B_1}_{magic} =\left(
\begin{array}{ccc}
 a & b & 0 \\
 b & 0 & a \\
 0 & a & b
\end{array}
\right),
\end{equation}
\begin{equation}
M^{B_2}_{magic} =\left(
\begin{array}{ccc}
 a & 0 & c \\
 0 & c & a \\
 c & a & 0
\end{array}
\right),
\end{equation}
\begin{equation}
M^{B_3}_{magic} =\left(
\begin{array}{ccc}
 a & 0 & c \\
 0 & 0 & a+c \\
 c & a+c & -c
\end{array}
\right),
\end{equation}
and
\begin{equation}
M^{B_4}_{magic} =\left(
\begin{array}{ccc}
 a & b & 0 \\
 b & -b & a+b \\
 0 & a+b & 0
\end{array}
\right).
\end{equation}
The magic mass matrices of type $B_1$ and $B_2$ are not allowed as
they predict $m_1=m_3$. The magic mass matrices of type 
$B_3$ and $B_4$ are not allowed because these textures
predict a very large value for the ratio 
$r=\Delta m^2_{12}/\Delta m^2_{23}$ when $\theta_{13}$ is small. 
We illustrate this tension between $r$ and $\theta_{13}$ 
for the magic mass matrices of type $B_3$ and $B_4$
in Section \ref{sec:pheno}.

\subsection{Class \textbf{C}}
The magic mass matrix of class \textbf{C} is
\begin{equation}
M^C_{magic} =\left(
\begin{array}{ccc}
 a & b & b \\
 b & 0 & a+b \\
 b & a+b & 0
\end{array}
\right).
\end{equation}
This mass matrix has $\mu-\tau$ symmetry and implies $\theta_{13}=0$.
Hence, it is not allowed. 

\section{Phenomenological implications} 
\label{sec:pheno}

The phenomenology of the textures $A_{1}$ and $A_{2}$ is related:
one can obtain the predictions for $A_{1}$ by making the transformations
\begin{equation}\label{eq:trans}
\theta_{23} \rightarrow \frac{\pi}{2}-\theta_{23}, \delta=\pi-\delta
\end{equation}
on the predictions of texture $A_{2}$. Hence, we study the 
phenomenological implications for texture $A_1$ only.
 
The above transformation [Eq. (\ref{eq:trans})] also relates 
the predictions for textures $B_3$ and $B_4$. So, we show 
the incompatibility of the magic mass matrix of type $B_3$ 
with the experimental data at the end of this section. 
Then, the Eq. (\ref{eq:trans}) automatically implies that the 
magic mass matrix of type $B_4$ is also inconsistent with 
the experimental data.

\subsection{Diagonalization of a magic mass matrix}
Any magic mass matrix $M$ can be diagonalized by 
a trimaximal mixing matrix $U=U_{TM}$ given in Eq. (\ref{eq:tm})
using the relation
\begin{equation}
U^{T}MU=M_{diag}
\end{equation}
where $M_{diag}$ is the diagonal mass matrix given by Eq. (\ref{eq:diag}).

The mixing angles can be calculated from $U$ using the relations:
\begin{equation}
s_{12}^{2}=\frac{|U_{12}|^{2}}{1-|U_{13}|^{2}}, s_{23}^{2}=\frac{|U_{23}|^{2}}{1-|U_{13}|^{2}} \textrm{ and } s_{13}^{2}=|U_{13}|^{2}.
\end{equation}
Substituting the elements of TM mixing matrix 
in the above equation, we get
\begin{equation}\label{eq:th12}
s_{12}^{2} = \frac{1}{3-2 \sin^2\theta},
\end{equation}
\begin{equation}\label{eq:th23}
s_{23}^{2}=\frac{1}{2} \left(1+\frac{\sqrt{3}   \sin 2 \theta \cos\phi}{3-2 \sin^2\theta}\right),
\end{equation}
and
\begin{equation}\label{eq:th13}
s_{13}^{2}=\frac{2}{3}\sin^2\theta.
\end{equation}

The CP violating phase $\delta$ can be calculated from the Jarlskog rephasing invariant measure of CP violation \cite{jarlskog}
\begin{equation}\label{eq:jcp}
J=Im(U_{11}U^*_{12}U^*_{21}U_{22})
\end{equation}  
using the relation
\begin{equation}\label{eq:jpar}
J=s_{12}s_{23}s_{13}c_{12}c_{23}c_{13}^2 \sin \delta.
\end{equation}
Substituting the elements of TM mixing matrix in Eq. (\ref{eq:jcp}),
we obtain
\begin{equation}\label{eq:jtm}
J=\frac{1}{6\sqrt{3}}\sin 2 \theta \cos \phi.
\end{equation}
From Eqs. (\ref{eq:jpar}) and (\ref{eq:jtm}), we get
\begin{equation}\label{eq:delta}
\csc^2 \delta = \csc ^2 \phi -\frac{3  
   \sin ^2 2 \theta \cot ^2\phi}{(3-2 \sin^2\theta)^2}.
\end{equation}

\subsection{Analysis of Class $A_1$}
We reconstruct the magic neutrino mass
matrix using the Eq. (\ref{eq:reco}) \textit{viz.}
\begin{equation}
M_{\nu}=U^*M_{diag}U^{\dagger}
\end{equation}
where $M_{\nu}=M_{magic}$ and $U=U_{TM}$.
To obtain the predictions for the neutrino mass matrix 
of the type $A_1$ given by Eq. (\ref{eq:a1}), we have to solve 
the two complex equations: $M_{\nu_{11}}=0$ and $M_{\nu_{12}}=0$. 

Solving the equation $M_{\nu_{11}}=0$, we get
\begin{equation}\label{eq:r12}
\frac{m_1}{m_2}= \frac{\sin 2 (\alpha -\beta )}{2 \sin2\beta\cos^2\theta} 
\end{equation}
and
\begin{equation}\label{eq:r23}
\frac{m_2}{m_3}= - \frac{2 \sin2 \beta \sin ^2\theta}{\sin 2 \alpha}. 
\end{equation}
Using these two equations, we evaluate
$m_1/m_3$ and invert the resulting relation to 
obtain 
\begin{equation}\label{eq:alpha}
\cot 2\alpha=\cot 2 \beta +\frac{m_1 }{m_3}\csc 2 \beta  \cot
   ^2 \theta.
\end{equation}

We note that the presence of a zero at (1,1)
entry in a magic mass matrix, through Eqs. 
(\ref{eq:r12}) and (\ref{eq:r23}), imply a beautiful 
sum-rule on neutrino masses:
\begin{equation}\label{eq:sumrule}
\frac{\sin 2 (\alpha -\beta
   )}{m_1}
   -\frac{2 \sin 2 \beta
   }{m_2}-\frac{\sin 2 \alpha
   }{m_3}=0.
\end{equation}
The texture zero at (1,1) entry in a magic mass matrix 
also gives a nice prediction for the ratio 
$r=\Delta m^2_{12}/\Delta m^2_{23}$. 
From Eqs. (\ref{eq:r12}) and (\ref{eq:r23}), we obtain
\begin{equation}\label{eq:rnu}
r = \frac{-\sin^2 2(\alpha - \beta)+4 \cos^2 \theta \sin^2 2 \beta}{\cot^2 \theta \sin^2 2 \alpha-4 \cos^2 \theta \sin^2 2 \beta}.
\end{equation}

\begin{figure*}[t]
\centering 
\includegraphics[scale=0.35]{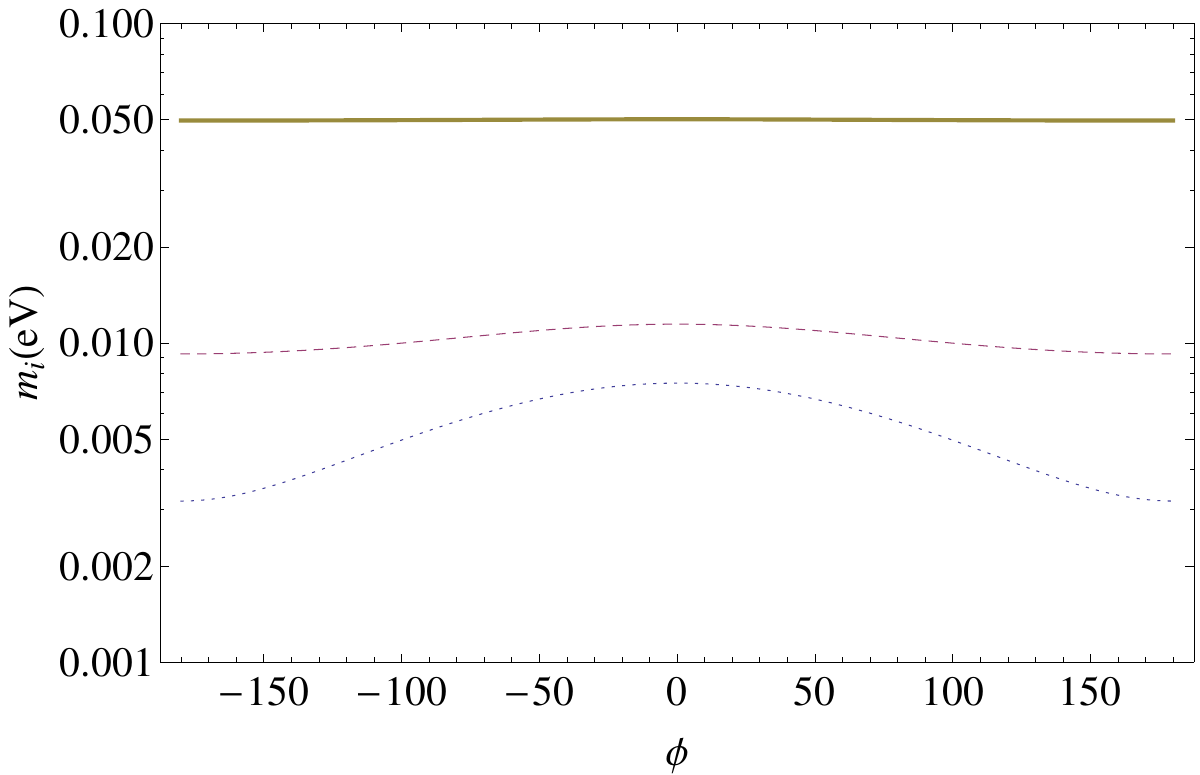}
\caption{The three neutrino masses $m_1$ (dotted line),
$m_2$ (dashed line) and $m_3$ (solid line) in $eV$
as functions of $\phi$ (in degrees).}
\label{fig:masses}
\end{figure*}

Instead of solving the second equation $M_{\nu_{12}}=0$, we solve
the equivalent complex equation $M_{\nu_{11}}=M_{\nu_{12}}$ by equating the real and imaginary parts of the two sides. After a little algebra, we obtain
\begin{equation}\label{eq:r13}
\frac{m_1}{m_3}=\frac{
   \sqrt{3}\sin 2 \beta \tan \theta +\sin (2 \beta -\phi )}{\sin \phi}
\end{equation}
and
\begin{equation}\label{eq:beta}
\tan 2 \beta =-\frac{\sqrt{3} \sin \phi}{\sqrt{3} \cos
   2 \theta \cos \phi+\sin 2 \theta}.
\end{equation}
Using Eq. (\ref{eq:r13}) to simplify Eq. (\ref{eq:alpha}), we obtain
\begin{equation}\label{eq:alphanew}
\cot 2\alpha=\cot
   \phi+\frac{\cot \theta \csc \phi}{\sqrt{3}}
\end{equation}

Equations (\ref{eq:beta}) and (\ref{eq:alphanew}) express the two
Majorana phases in terms of the two TM parameters ($\theta$
and $\phi$). Substituting these two equation in Eq. (\ref{eq:rnu}), 
we obtain the most important result of this work as
\begin{equation}\label{eq:r}
r=\tan^2\theta.
\end{equation}
It is interesting that $r$ comes out to be independent of the phase 
$\phi$.

\begin{figure*}[t]
\centering 
{\includegraphics[scale=0.4]{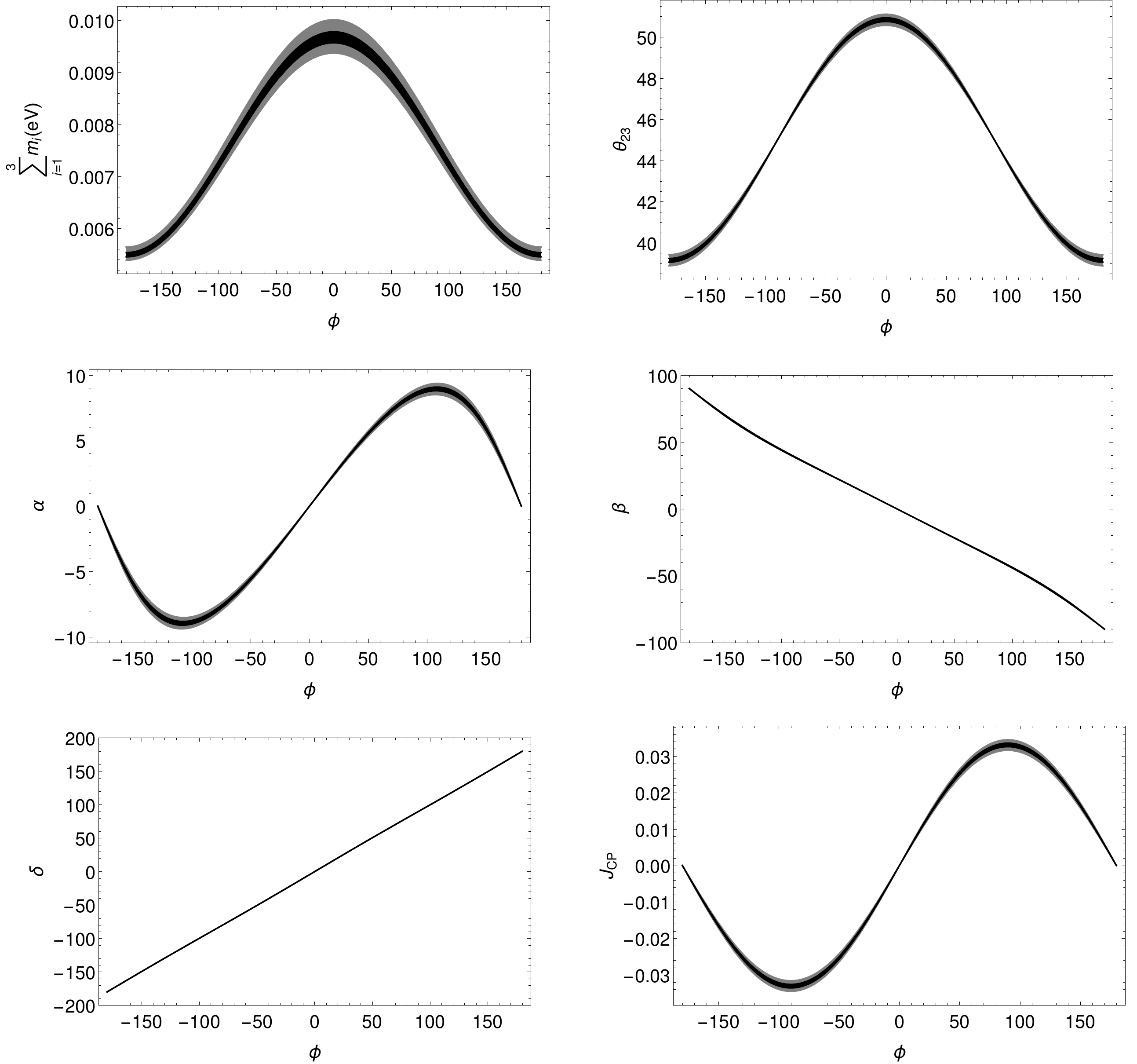}} 
\caption{The neutrino parameters 
$\displaystyle{\sum_{i=1}^{3}} m_i$, $\theta_{23}$, 
$\alpha$, $\beta$, $\delta$, and $J_{CP}$
as functions of $\phi$. All phases and angles are in degrees. The dark (gray)
bands depict the $1\sigma$ ($3\sigma$) allowed regions.}
\label{fig:plots}
\end{figure*}

We also substitute Eqs. (\ref{eq:beta}) and (\ref{eq:alphanew}) in the
three mass ratios given by 
Eqs. (\ref{eq:r12}), (\ref{eq:r13}), and (\ref{eq:r23}) to calculate the three
neutrino masses. Finally, we express $\theta$ in terms of $r$ everywhere
using Eq. (\ref{eq:r}). Hence, we can express the three neutrino masses 
in terms of the three parameters:
$\Delta m^{2}_{12}$, $r$ and $\phi$. 
We obtain 
\begin{equation}
m_{1}=\sqrt{\Delta m^{2}_{12}} \sqrt{\frac{1+3r
+2\sqrt{3} \sqrt{r} \cos \phi}{3-3r-2
   \sqrt{3} \sqrt{r} \cos \phi}},
\end{equation}
\begin{equation}
m_{2}=\frac{2 \sqrt{\Delta m^{2}_{12}}}{\sqrt{3-3r-2 \sqrt{3}
   \sqrt{r} \cos \phi}},
\end{equation}
and
\begin{equation}
m_{3}=\sqrt{\Delta m^{2}_{12}}\sqrt{
\frac{3+r-2 \sqrt{3} \sqrt{r} \cos \phi}
{3-3r-2 \sqrt{3} \sqrt{r} \cos\phi}}.
\end{equation} 
Now, we can use the experimental data \cite{data}
for $\Delta m^{2}_{12}$ and $\Delta m^{2}_{23}$.
Since, $\Delta m^{2}_{12}=(7.50 \pm 0.18) \times 10^{-5}eV^2$
and $r=(3.149 \pm 0.098) \times 10^{-2}$ \cite{data}, 
the three masses are essentially functions 
of the phase $\phi$ (Fig. \ref{fig:masses}). 
We also depict the sum of the three neutrino masses
$\displaystyle{\sum_{i=1}^{3}} m_i$ as a function of $\phi$ 
in Fig. \ref{fig:plots}.

The three mixing angles, calculated from Eqs. 
(\ref{eq:th12}-\ref{eq:th13}), are
\begin{equation}
\sin^{2}\theta_{12}=\frac{1+r}{3+r},
\end{equation}
\begin{equation}
\sin^{2}\theta_{23}=\frac{1}{2}+\frac{\sqrt{3} \sqrt{r} \cos\phi}{r+3}
\end{equation}
and
\begin{equation}
\sin^{2}\theta_{13}=\frac{2 r}{3 (r+1)}.
\end{equation}
The two mixing angles $\theta_{12}$ and $\theta_{13}$ 
are functions of $r$ only. Substituting the value of
$r$, we obtain $\theta_{12}=35.67^o \pm 0.01^o$ and 
$\theta_{13}=8.20^o \pm 0.12^o$. For comparison,
the experimental values are $\theta_{12}=33.48^o \pm 0.78^o$ 
and $\theta_{13}=8.50^o \pm 0.21^o$. The predicted and experimental 
values of $\theta_{12}$ become compatible at about $2.8\sigma$ C.L. 
This discrepancy is, however, a generic feature of TM mixing. 
One possible way to diffuse this tension with the data is to 
consider charged lepton corrections. We have presented our textures in a basis 
in which the charge lepton mass matrix is diagonal and the effective neutrino 
mass matrix is magic with two zeros. However, in a model realization of these textures, 
the charged lepton mass matrix can have small off-diagonal terms that will 
give corrections to the neutrino mixing angles. One can arrange
these corrections to bring $\theta_{12}$ to its experimental value 
while keeping other two angles within the allowed ranges.

The mixing angle $\theta_{23}$ is 
a function of the phase $\phi$ after substituting for $r$. We depict 
the mixing angle $\theta_{23}$ as the function of phase $\phi$ 
in Fig. \ref{fig:plots}.

We can calculate the three CP violating phases from Eqs. (\ref{eq:alphanew}), (\ref{eq:beta}), and (\ref{eq:delta}). We obtain
\begin{equation}
\cot 2\alpha=\cot \phi+\frac{\csc \phi}{\sqrt{3}
   \sqrt{r}},
\end{equation}
\begin{equation}
\tan 2\beta=-\frac{\sqrt{3} (1+r) \sin \phi}{2 \sqrt{r}+
\sqrt{3}(1-r) \cos \phi},
\end{equation}
and
\begin{equation}
\tan\delta=\frac{3+r}{3-r}\tan\phi.
\end{equation}
The Jarlskog invariant $J$, calculated from Eq. (\ref{eq:jtm}), is
\begin{equation}
J=\frac{\sqrt{r} \sin \phi}{3\sqrt{3} (1+r)}.
\end{equation}
The three CP violating phases 
($\alpha$, $\beta$, and $\delta$) 
depend upon the ratio $r$ and the unknown phase $\phi$.
Therefore, we can plot $\alpha$, 
$\beta$, $\delta$, and $J$ as functions of $\phi$ 
by just plugging in one experimental number 
$r$ (Fig. \ref{fig:plots}). 

This high level of predictability makes these textures good
candidates for model-building. It is rarely
seen that a neutrino mass model can predict the nine neutrino 
parameters using just two inputs from the experiments: 
$\Delta m^{2}_{12}$ and $\Delta m^{2}_{23}$.
We present an $A_4$ based model for these two textures
in the next section.

\begin{figure*}[t]
\centering 
\includegraphics[scale=0.35]{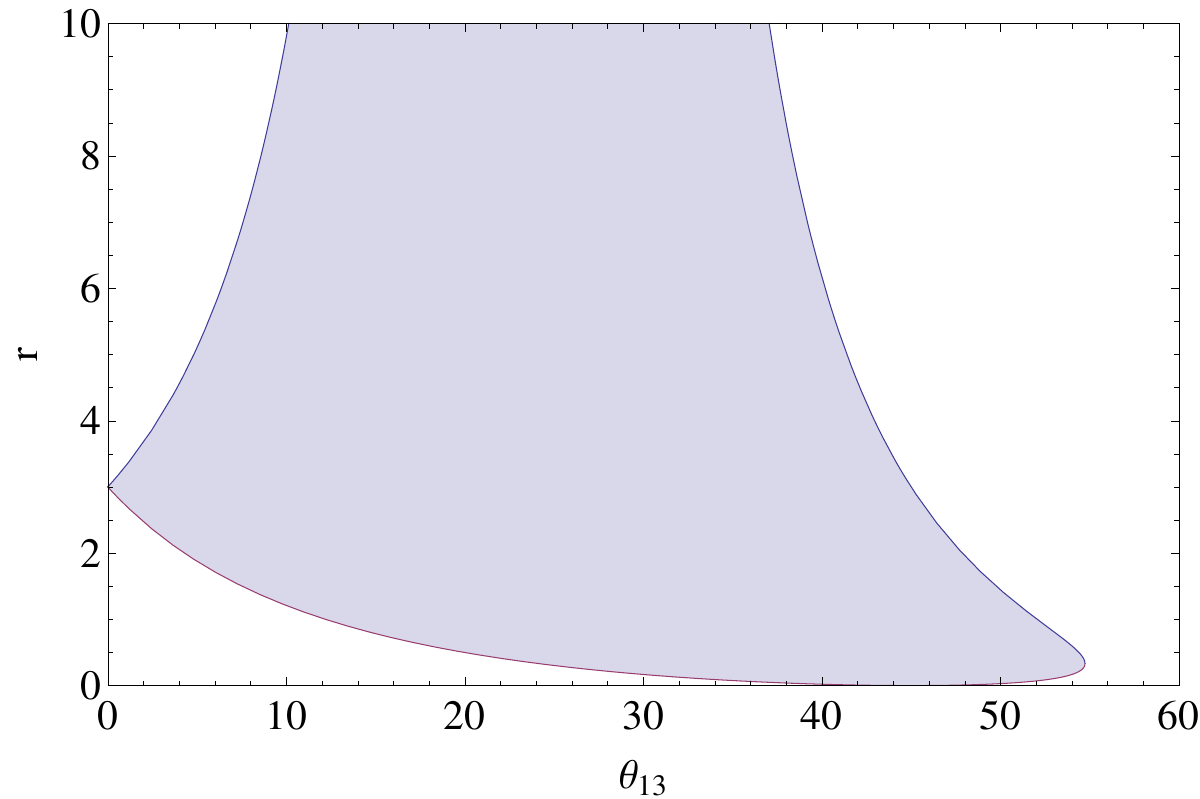}
\caption{The ratio $r=\Delta m^2_{12}/\Delta m^2_{23}$ as a function of 
$\theta_{13}$ (degrees) for magic mass matrix of type $B_3$.}
\label{fig:b3}
\end{figure*} 

\subsection{Inconsistency of Class $B_3$}
The magic mass matrix of type $B_3$ has zeros at $(1,2)$ and $(2,2)$
entries. This implies following two complex equations:

\begin{equation}
\frac{m_1}{m_2}e^{2 i \alpha}=
\frac{2 \left(\sqrt{3} e^{-i \phi} \sin
   ^2 \theta+\sqrt{3} e^{i \phi} \cos ^2 \theta+2
   \sin 2 \theta\right)}{\left(1-3
   e^{2 i \phi}\right) \sin 2 \theta+2
   \sqrt{3} e^{i \phi} \cos 2 \theta}
\end{equation}
and 
\begin{equation}
\frac{m_2}{m_3}e^{2 i \beta}=\frac{\sqrt{3}+3 e^{i \phi} \cot
   \theta}{\sqrt{3}-3 e^{-i \phi} \cot \theta}.
\end{equation}
Using absolute squares of these ratios, we can calculate the ratio $r$ as
\begin{equation}
r=\frac{1- \left| \frac{m_1}{m_2}e^{2 i \alpha} \right|^2}{\left( 
\left| \frac{m_2}{m_3}e^{2 i \beta} \right| \right)^{-1}-1}.
\end{equation}
Using these expressions, we express $r$ as a
function of $\theta_{13}$ \mbox{(Fig. \ref{fig:b3})}
by substituting the value of $\theta$ in terms of $\theta_{13}$ from 
Eq. (\ref{eq:th13}).
We find that $r$ has a minimum value $r=0$ at the point
$(\theta_{13}=\pi/4,\phi=\pi)$. We obtain the experimental value of
$r$ only in a small interval around this point for 
$\theta_{13} \in [40^{\circ},50^{\circ}]$. As $\theta_{13}$ decreases, the 
minimum value of $r$ increases. 
It is clear that we cannot have both $r$ and $\theta_{13}$
in their experimentally allowed ranges simultaneously. Hence, this texture
is inconsistent with the experimental data.

\section{The $A_4$ Model}
\label{sec:model}

We present an $A_4$ model in the framework of 
type-I+II see-saw mechanism \cite{seesaw1, seesaw2} to obtain the neutrino mass matrices
studied in this work. Apart from the three left-handed lepton doublets $D_{l_L}$ and
three right-handed charged leptons $l_R$ 
(where $l=e,\mu~ \text{and}~ \tau$), 
we introduce six $SU(2)_L$ doublet Higgs
fields $\psi_i$ and $\varphi_i$, 
(where $i=1,2~ \text{and}~ 3$) and 
a $SU(2)_L$ triplet Higgs field $\Delta$.
We depict the transformation properties of the fields 
present in our model in Table \ref{tab:trans}. 
In addition to $A_4$ symmetry, we also need 
a $Z_2$ symmetry to prevent the coupling 
of the charged leptons (neutrinos) with scalars $\varphi_i$ 
($\psi_i$). These transformation properties lead to the following 
Lagrangian for the leptons that is invariant under $A_4$ and $Z_2$.
\begin{eqnarray}
-\mathcal{L} & = & y_1 (\overline{D}_{e_L} \psi_1 + \overline{D}_{\mu_L} \psi_2 
+ \overline{D}_{\tau_L} \psi_3)_{\underline{1}} e_{R_{1}}  \nonumber \\
& + & y_2 (\overline{D}_{e_L} \psi_1 + \omega^2 
\overline{D}_{\mu_L} \psi_2 +  \omega \overline{D}_{\tau_L} \psi_3)_{\underline{1}^{\prime \prime}} \tau_{R_{1^{\prime}}}  \nonumber \\
& + & y_3 (\overline{D}_{e_L} \psi_1 + 
\omega \overline{D}_{\mu_L} \psi_2 +  
\omega^2 \overline{D}_{\tau_L} \psi_3)_{\underline{1}^{\prime}} \mu_{R_{1^{\prime \prime}}} \nonumber \\
& + & y_4 (\overline{D}_{e_L} \tilde{\varphi}_1 + \overline{D}_{\mu_L} \tilde{\varphi}_2 + \overline{D}_{\tau_L} \tilde{\varphi}_3)_{\underline{1}} \nu_{R_1}  \nonumber \\ 
& - & y_{\Delta} (D_{e_L}^T C^{-1} D_{e_L} + 
\omega^2 D_{\mu_L}^T C^{-1} D_{\mu_L}  \nonumber \\ 
& + &
\omega D_{\tau_L}^T C^{-1} D_{\tau_{L}})_{\underline{1}^{\prime \prime}} i \tau_2 \Delta_{1^\prime} \nonumber \\ 
& - & m_R( \nu_{R}^T C^{-1} \nu_{R}) + \ \ \textrm{h.c.}
\end{eqnarray}
where $\tilde{\varphi}=i \tau_2 \varphi^*$.
 
We assume the following vacuum expectation values (vevs) of the
Higgs fields: $\langle \psi \rangle_o = v_{\psi}(1, 1, 1)^T$ which leads to the charged lepton mass matrix
\begin{equation}
 m_l = \left(
\begin{array}{ccc}
y_1 v_{\psi} & y_2 v_{\psi} & y_3 v_{\psi} \\ 
y_1 v_{\psi} & y_2 \omega v_{\psi} & y_3 \omega^2 v_{\psi} \\
y_1 v_{\psi} & y_2 \omega^2 v_{\psi} & y_3 \omega v_{\psi}
\end{array}
\right).
 \end{equation}
For the type-I see-saw contribution, we assume that $\varphi_i$ develop vevs along the direction $\langle \varphi \rangle_o = v_{\varphi}(0, -1, 1)^T$. Such a vacuum alignment has been obtained in references \cite{vev} for $SU(2)_L$ and $A_4$ triplet scalars by allowing specific terms in the scalar potential which break $A_4$ softly. This choice of vevs leads to the following Dirac neutrino mass matrix 
 \begin{equation}
 m_D = y_4 v_{\varphi} (0,-1,1)^T.
 \end{equation}
We have only one right handed neutrino with mass $m_R$.
Using the type-I see-saw mechanism, the effective neutrino mass matrix is
$m^I_{\nu} \approx m_D m_R^{-1} m_D^T$,
\begin{equation}
 m^I_{\nu} = c \left(
\begin{array}{ccc}
0 & 0 & 0 \\ 0 & 1 & -1 \\ 0& -1 & 1
\end{array}
\right)
\end{equation}
where $c=y_4^2 v_{\psi}^2/m_R$.
When the $SU(2)_L$ triplet Higgs acquires a non-zero and small vev,
we get the following type-II see-saw contribution to the effective neutrino mass matrix:
 \begin{equation}
 m^{II}_{\nu} = \Delta \left(
\begin{array}{ccc}
1 & 0 & 0 \\ 0 & \omega^2 & 0 \\ 0& 0 & \omega
\end{array}
\right)
\end{equation}
where $\Delta=y_{\Delta} v_{\Delta}$.
The combined effective neutrino mass matrix $m_{\nu}=m^{I}_{\nu}+m^{II}_{\nu}$ 
from type-I+II see-saw mechanism becomes 
\begin{equation}
 m_\nu = \left(
\begin{array}{ccc}
\Delta & 0 & 0 \\ 0 & c + \omega^2 \Delta & -c \\ 0 & -c & c+ \omega \Delta
\end{array}
\right).
\end{equation}

In the symmetry basis, the charged lepton mass matrix $m_l$
is not diagonal. We make a transformation to the basis
where the charge lepton mass matrix is diagonal with the
transformation $M_{l}=U_L^{\dagger} m_l U_R$, where
\begin{equation}
 U_L=\frac{1}{3}\left(
\begin{array}{ccc}
1 & 1 & 1 \\ 1 & \omega & \omega^2 \\ 1 & \omega^2 & \omega
\end{array}
\right),
\end{equation}
and $U_R$ is a unit matrix.
In this basis where $M_l$ is diagonal, the effective neutrino mass
matrix becomes:
\begin{equation}
 M_\nu = \left(
\begin{array}{ccc}
0 & 0 & \Delta \\ 
0 & \Delta-c & c \\ 
\Delta & c & -c
\end{array}
\right).
\end{equation}
This is the mass matrix of type $A_1$ having magic symmetry and
two texture zeros.

A similar mechanism with $SU(2)_L$ triplet Higgs $\Delta$ transforming
as $1^{\prime \prime}$ instead of $1^{\prime}$ will give 
the neutrino mass matrix:
\begin{equation}
 M_\nu = \left(
\begin{array}{ccc}
0 & b & 0 \\ b & -a & a\\ 0 & a & b-a 
\end{array}
\right).
\end{equation}
This is the mass matrix of type $A_2$ having magic symmetry and
two texture zeros.
  
\begin{table}[!]
\begin{center}
\begin{tabular}{ccccccccc}
\hline 
Fields & $D_{l_{L}}$ & $l_R $ & $\nu_{R}$ & $\psi $ & $\varphi $ & $\Delta$ \\ 
\hline 
$SU(2)_L$ & 2 & 1 & 1 & 2 & 2 & 3  \\ 
$A_4$ & 3 & $1,1^{\prime \prime},1^\prime$ & 1 & 3 & 3 & $1^{\prime}$ \\ 
$Z_2$ & 1 & 1 & -1 & 1 & -1 & 1 \\ 
\hline 
\end{tabular}
\end{center}
\caption{Transformation properties of various fields: $D_{l_{L}} ~ (D_{e_{L}}, D_{\mu_{L}}, D_{\tau_{L}})^T$, 
$l_R ~ (e_R, \mu_R, \tau_R)^{T}$, $\nu_{l_{R}}$,  $\psi ~ (\psi_1, \psi_2, \psi_3)^T$, 
$\varphi (\varphi_1, \varphi_2, \varphi_3)^T$ and $\Delta$.}
\label{tab:trans}
\end{table} 

Our model requires 6 Higgs doublets, 3 of which couple to charged leptons 
\mbox{[Table II].} In such multi Higgs models, the flavor changing neutral 
currents can contribute to charged lepton flavor violating decays. However, 
an explicit calculation is beyond the scope of present work due to the complexity 
of Higgs sector of our model. Nevertheless, there exist models in literature e.g. 
Ref. \cite{ma} where the charged lepton Yukawa Lagrangian 
(including the $A_4$ assignments of charged lepton and scalar fields) are
similar to our model. The flavor violating decays 
of leptons for our model can be studied in a manner similar to Ref. \cite{ma}.

\section{Conclusions}
\label{sec:conclusions}
We study the phenomenological implications of two texture zeros in the 
magic neutrino mass matrix. In absence of magic symmetry,
there are seven allowed patterns for the presence of two zeros in the 
neutrino mass matrix. The additional constraint of magic symmetry 
disallows five of these patterns. The two allowed patterns 
are of the type $A_1$ and $A_2$. 
The combination of magic symmetry and texture zeros make these classes
very predictive. We can express all the nine neutrino observables 
(the three masses, the three mixing angles, and the three CP violating
phases) as the function of $\phi$ by plugging in just two experimental
parameters ($\Delta m^{2}_{12}$ and $\Delta m^{2}_{23}$). In particular,
$\theta_{12}$ and $\theta_{13}$ do not even depend upon
the phase $\phi$ and can be expressed as functions of the ratio 
$r = \Delta m^{2}_{12}/\Delta m^{2}_{23}$ as 
$\sin^{2}\theta_{12}=\frac{1+r}{3+r}$ and 
$\sin^{2}\theta_{13}=\frac{2 r}{3 (r+1)}$. Finally, we have derived
these highly predictive mass matrices from a neutrino mass model based upon
the symmetry group $A_4$. 

\acknowledgements{R. R. G. acknowledges the  financial support from 
Department of Science and Technology, Government of India under the 
Grant No. SB/FTP/PS-128/2013. S.K acknowledges the financial support from
Department of Science and Technology, Government of India, under the Grant 
No. SR/FTP/PS-123/2011; and from University of Delhi, under the Research and 
Development Grant No. RC/2015/9677.
}

\renewcommand{\theequation}{A-\arabic{equation}}
\setcounter{equation}{0}
\section*{A \ \ The Group $A_4$} 
$A_4$ is the group of even permutations of four objects having twelve elements. Geometrically, it can be viewed as the group of rotational symmetries of the tetrahedron. $A_4$ has four inequivalent irreducible representations (IRs) which are three singlets \textbf{1}, $\textbf{1}^\prime$ and $\textbf{1}^{\prime\prime}$, and one triplet \textbf{3}. The group $A_4$ is generated by two generators $S$ and $T$ such that 
\begin{equation}
S^2 = T^3 = (S T)^3 = 1.
\end{equation}
The one dimensional unitary IRs are
\begin{equation}
\textbf{1} \  S = 1 \ \ T = 1, \ \
\textbf{1}^{\prime} \  S = 1 \ \ T = \omega, 
\textbf{1}^{\prime\prime} \  S = 1 \ \ T = \omega^2.
\end{equation}
The three dimensional unitary IR is 
 \begin{equation}
 S = \left(
\begin{array}{ccc}
1 & 0 & 0 \\ 0 & -1 & 0 \\ 0& 0 & -1
\end{array}
\right), \ \ T = \left(
\begin{array}{ccc}
0 & 1 & 0 \\ 0 & 0 & 1 \\ 1& 0 & 0
\end{array}
\right).
\end{equation}
The multiplication rules of the IRs are as follows
\begin{equation}
\textbf{1}^\prime \otimes \textbf{1}^\prime = \textbf{1}^{\prime \prime}, \ \textbf{1}^{\prime \prime} \otimes \textbf{1}^{\prime \prime} = \textbf{1}^{\prime}, \ \textbf{1}^{\prime} \otimes \textbf{1}^{\prime \prime} = \textbf{1}.
\end{equation}
The product of two $\textbf{3}$'s gives 
\begin{equation}
\textbf{3} \otimes \textbf{3} = \textbf{1} \oplus \textbf{1}^\prime \oplus \textbf{1}^{\prime \prime} \oplus \textbf{3}_s \oplus \textbf{3}_a,
\end{equation}
where $s$($a$) denotes the symmetric(anti-symmetric) product. 
Let $(x_1, x_2, x_3)$ and $(y_1, y_2, y_3)$ denote 
the basis vectors of two $\textbf{3}$'s. Then the IRs obtained from 
their products are
\begin{align}
(\textbf{3} \otimes \textbf{3})_{\textbf{1}} & = x_1 y_1 + x_2 y_2 + x_3 y_3 \\
(\textbf{3} \otimes \textbf{3})_{\textbf{1}^\prime} & = x_1 y_1 + \omega x_2 y_2 + \omega^2 x_3 y_3 \\
(\textbf{3} \otimes \textbf{3})_{\textbf{1}^{\prime \prime}} & = x_1 y_1 + \omega^2 x_2 y_2 + \omega x_3 y_3 \\
(\textbf{3} \otimes \textbf{3})_{\textbf{3}_s} & = (x_2 y_3 + x_3 y_2, x_3 y_1 + x_1 y_3, x_1 y_2 + x_2 y_1) \\
(\textbf{3} \otimes \textbf{3})_{\textbf{3}_a} & = (x_2 y_3 - x_3 y_2, x_3 y_1 - x_1 y_3, x_1 y_2 - x_2 y_1).
\end{align}


\begin{thebibliography}{99}

\bibitem{th13} P. Adamson et al. [MINOS Collaboration], \textit{Phys. Rev. Lett.} \textbf{107}, 181802 (2011), [arXiv:1108.0015 [hep-ex]]; Y. Abe et al., [Double Chooz Collaboration], \textit{Phys. Rev. Lett.} \textbf{108}, 131801 (2012), [arXiv:1112.6353 [hep-ex]]; F. P. An et al., [Daya Bay Collaboration], \textit{Phys. Rev. Lett.} \textbf{108}, 171803 (2012), [arXiv:1203.1669 [hep-ex]]; Soo-Bong Kim, for RENO Collaboration, \textit{Phys. Rev. Lett.} \textbf{108}, 191802 (2012), [arXiv:1204.0626 [hep-ex]].

\bibitem{mutau} T.~Fukuyama and H.~Nishiura, [hep-ph/9702253]; R.~N.~Mohapatra and S.~Nussinov, Phys.\ Rev.\ D {\bf 60}, 013002 (1999), [hep-ph/9809415]; K.~R.~S.~Balaji, W.~Grimus and T.~Schwetz, Phys.\ Lett.\ B {\bf 508}, 301 (2001), [hep-ph/0104035]; C.~S.~Lam, Phys.\ Lett.\ B {\bf 507}, 214 (2001), [hep-ph/0104116]; W.~Grimus and L.~Lavoura, JHEP {\bf 0107}, 045 (2001), [hep-ph/0105212].

\bibitem{hps} P. F. Harrison, D. H. Perkins and W. G. Scott, \textit{Phys. Lett.} \textbf{B 530}, 167 (2002), [hep-ph/0202074]; Zhi-zhong Xing, \textit{Phys. Lett.} \textbf{B 533}, 85 (2002), [hep-ph/0204049]; P. F. Harrison and W. G. Scott, \textit{Phys. Lett.} \textbf{B 535}, 163 (2002), [hep-ph/0203209].

\bibitem{magic} C.~S.~Lam, Phys.\ Lett.\ B {\bf 640} (2006) 260, [hep-ph/0606220]; P.~F.~Harrison and W.~G.~Scott, Phys.\ Lett.\ B {\bf 594}, 324 (2004), [hep-ph/0403278]; R.~Friedberg and T.~D.~Lee, HEPNP {\bf 30}, 591 (2006), [hep-ph/0606071].

\bibitem{tm2} J.~D.~Bjorken, P.~F.~Harrison and W.~G.~Scott, Phys.\ Rev.\ D {\bf 74}, 073012 (2006), [hep-ph/0511201]; X.~G.~He and A.~Zee, Phys.\ Lett.\ B {\bf 645}, 427 (2007), [hep-ph/0607163]; Carl H. Albright, Werner Rodejohann Eur.Phys.J. C62 (2009) 599-608 [arXiv:0812.0436 [hep-ph]]; Carl H. Albright, Alexander Dueck, Werner Rodejohann Eur.Phys.J. C70 (2010) 1099-1110, [arXiv:1004.2798 [hep-ph]]; X.~G.~He and A.~Zee, Phys.\ Rev.\ D {\bf 84}, 053004 (2011), [arXiv:1106.4359 [hep-ph]]; Sanjeev Kumar, Phys.Rev.D82 (2010) 013010, [arxiv:1007.0808 [hep-ph]]; ibid 88 (2013) 1, 016009, [arXiv:1305.0692 [hep-ph]].

\bibitem{tm2models} N.~Haba, A.~Watanabe and K.~Yoshioka, Phys.\ Rev.\ Lett.\  {\bf 97}, 041601 (2006), [hep-ph/0603116]; W.~Grimus and L.~Lavoura, JHEP {\bf 0809}, 106 (2008), [arXiv:0809.0226 [hep-ph]]; H.~Ishimori, Y.~Shimizu, M.~Tanimoto and A.~Watanabe, Phys.\ Rev.\ D {\bf 83}, 033004 (2011), [arXiv:1010.3805 [hep-ph]]; Y.~Shimizu, M.~Tanimoto and A.~Watanabe, Prog.\ Theor.\ Phys.\  {\bf 126}, 81 (2011), [arXiv:1105.2929 [hep-ph]]; S.~F.~King and C.~Luhn, JHEP {\bf 1109}, 042 (2011), [arXiv:1107.5332[hep-ph]]; S. Dev, S. Gupta, R. R. Gautam, Phys.Lett. B702 (2011) 28-33, [arXiv:1106.3873 [hep-ph]]; S. Dev, R. R. Gautam and L. Singh, Phys.Lett. B708 (2012) 284-289, [arXiv:1201.3755 [hep-ph]]; 


\bibitem{fgm} Paul H. Frampton, Sheldon L. Glashow and Danny Marfatia, \textit{Phys. Lett.} \textbf{B 536}, 79 (2002), [hep-ph/0201008].

\bibitem{xingtz} H. Fritzsch, Zhi-zhong Xing, S. Zhou, \textit{JHEP } \textbf{1109}, 083 (2011), [arXiv:1108.4534 [hep-ph]].

\bibitem{tz} Zhi-zhong Xing, \textit{Phys. Lett.} \textbf{B 530}, 159 (2002), [hep-ph/0201151]; Bipin R. Desai, D. P. Roy and Alexander R. Vaucher, \textit{Mod. Phys. Lett.} \textbf{A 18}, 1355 (2003), [hep-ph/0209035]; A. Merle, W. Rodejohann, \textit{Phys. Rev} \textbf{D 73}, 073012 (2006), [hep-ph/0603111];  
S. Dev, Sanjeev Kumar, S. Verma and S. Gupta, \textit{Nucl. Phys.} \textbf{B 784}, 103 (2007), [hep-ph/0611313]; 
S. Dev, S. Kumar, S. Verma and S. Gupta, \textit{Phys. Rev.} \textbf{D 76}, 013002 (2007), [hep-ph/0612102]; 
G. Ahuja, S. Kumar, M. Randhawa, M. Gupta, S. Dev, \textit{Phys. Rev.} \textbf{D 76}, 013006 (2007), [hep-ph/0703005]; 
S. Kumar, \textit{Phys. Rev.} \textbf{D 84}, 077301 (2011), [arXiv:1108.2137 [hep-ph]]; 
S. Dev, S. Kumar, S. Verma, \textit{Phys. Rev.} \textbf{D 79}, 033001 (2009), [hep-ph/0612102]; 
P. O. Ludl, S. Morisi, E. Peinado, \textit{Nucl. Phys.} \textbf{B 857}, 411 (2012), [arXiv:1109.3393 [hep-ph]]; 
D. Meloni, G. Blankenburg, \textit{Nucl. Phys.} \textbf{B 867}, 749 (2013), [arXiv:1204.2706 [hep-ph]]; 
W. Grimus, P. O. Ludl, \textit{J. Phys.} \textbf{G 40}, 055003 (2013) [arXiv:1208.4515 [hep-ph]]; 
J. Liao, D. Marfatia, K. Whisnant, [arXiv:1311.2639 [hep-ph]]; D. Meloni, A. Meroni, E. Peinado, \emph{Phys. Rev.} {\bf D 89} (2014) 053009, [arXiv:1401.3207 [hep-ph]]; 
S.~Dev, R.~R.~Gautam, L.~Singh and M.~Gupta, Phys.\ Rev.\ D {\bf 90}, no. 1, 013021 (2014), [arXiv:1405.0566 [hep-ph]]; 
G.~Ahuja, S.~Sharma, P.~Fakay and M.~Gupta,  Mod.\ Phys.\ Lett.\ A {\bf 30}, 1530025 (2015), [arXiv:1604.03339 [hep-ph]].

\bibitem{s4} C.~S.~Lam, Phys.\ Rev.\ D {\bf 78}, 073015 (2008), [arXiv:0809.1185 [hep-ph]]; C.~S.~Lam, [arXiv:0907.2206 [hep-ph]].

\bibitem{grimus} W.~Grimus, L.~Lavoura and P.~O.~Ludl, J.\ Phys.\ G {\bf 36}, 115007 (2009), [arXiv:0906.2689 [hep-ph]]

\bibitem{king} S.~F.~King and C.~Luhn, JHEP {\bf 0910}, 093 (2009), [arXiv:0908.1897 [hep-ph]].

\bibitem{data} M. C. Gonzalez-Garcia et al., Nuclear Physics \textbf{B 908}, 199 (2016), [arXiv:1512.06856 [hep-ph]].

\bibitem{jarlskog} C. Jarlskog, \textit{Phys. Rev. Lett.} \textbf{55}, 1039 (1985).

\bibitem{seesaw1} P. Minkowski, \textit{Phys. Lett.} \textbf{B 67}, 421 (1977); T. Yanagida, \textit{Proceedings of the Workshop on the Unified Theory and the Baryon Number in the Universe} (O. Sawada and A. Sugamoto, eds.), KEK, Tsukuba, Japan, 1979, p. 95: M. Gell-Mann, P. Ramond, and R. Slansky, \textit{Complex spinors and unified theories in supergravity} (P. Van Nieuwenhuizen and D. Z. Freedman, eds.), North Holland, Amsterdam, 1979, p.315; R. N. Mohapatra and G. Senjanovic, \textit{Phys. Rev. Lett.} \textbf{44}, 912 (1980).

\bibitem{seesaw2} W. Konetschny and W. Kummer, \textit{Phys. Lett.} \textbf{B 70}, 433 (1977); T. P. Cheng and L. F. Li, \textit{Phys. Rev.} \textbf{D 22}, 2860 (1980); J. Schechter and J. W. F. Valle, \textit{Phys. Rev.} \textbf{D 22}, 2227 (1980); G. Lazarides Q. Shafi and C. Wetterich, \textit{Nucl. Phys.} \textbf{B 181}, 287 (1981); R. N. Mohapatra and G. Senjanovic, \textit{Phys. Rev.} \textbf{D 23}, 165 (1981).

\bibitem{vev} S.~Gupta, A.~S.~Joshipura and K.~M.~Patel, Phys.\ Rev.\ D {\bf 85}, 031903 (2012), [arXiv:1112.6113 [hep-ph]]; E.~Ma, Phys.\ Rev.\ D {\bf 70}, 031901 (2004), [hep-ph/0404199]; E.~Ma and D.~Wegman, Phys.\ Rev.\ Lett.\  {\bf 107}, 061803 (2011), [arXiv:1106.4269 [hep-ph]]. 

\bibitem{ma} E.~Ma and G.~Rajasekaran, Phys.\ Rev.\ D {\bf 64}, 113012 (2001), [hep-ph/0106291].
 
\end{thebibliography}
\end{document}